\documentclass[cameraready]{Interspeech}

\title{A study on weakly-supervised training approaches for phoneme-level pronunciation scoring}

\author[affiliation={1,2}, orcid=0009-0008-1306-7761, correspondingauthor]{Jazmín}{Vidal}
\author[affiliation={2}, orcid=0000-0002-0426-8683]{Luciana}{Ferrer}

\address{
    $^1$ Departamento de Computaci\'{o}n, FCEyN, Universidad de Buenos Aires (UBA), Argentina \\
    $^2$ Instituto de Investigaci\'{o}n en Ciencias de la Computaci\'{o}n (ICC), CONICET-UBA, Argentina
}

\email{jvidal@dc.uba.ar, lferrer@dc.uba.ar}

\keywords{computer assisted pronunciation training, pronunciation scoring, phoneme level, weak supervision}

\usepackage{comment}
\usepackage{multirow}

\begin{document}

\maketitle

\begin{abstract}
Phoneme-level computer-assisted pronunciation training systems typically rely on phoneme-level annotations, which are costly and scarce. In this work, we investigate whether phoneme-level mispronunciation information can be learned without phoneme-level supervision by exploiting higher-level pronunciation labels. Specifically, we study a weakly supervised setting in which models are trained using only utterance- or word-level pronunciation labels and analyze whether this supervision induces useful phoneme-level score predictions. We further consider a two-stage training scenario in which a model trained only with utterance-level labels is finetuned using a limited number of carefully-selected phoneme-level labeled utterances. We find that, using our proposed architecture and selection process, the two-stage process leads to comparable results to those obtained with full phoneme-level supervision, requiring only a small fraction of phoneme-level labels.
\end{abstract}

\section{Introduction}
\label{sec:intro}
Computer-assisted pronunciation training (CAPT) systems aim to detect and provide feedback on multiple aspects of L2 speech, including segmental and suprasegmental accuracy, fluency, and overall proficiency. These aspects can be evaluated at different granularities (phoneme, word, or sentence) and modeled separately or jointly \cite{witt2012automatic, elke2023automatic}. CAPT systems that provide feedback at phoneme level have shown to improve learning and motivation \cite{neri2008effectiveness}.

Most phoneme-level mispronunciation detection systems are trained with supervision, relying on manually labeled non-native data. This includes approaches that train classifiers to distinguish correct from incorrect realizations \cite{strik2009comparing} as well as methods that recognize non-native phoneme sequences and compare them to canonical targets \cite{leung2019cnn}. To make efficient use of limited annotated data, recent works have adopted transfer learning approaches based on automatic speech recognition (ASR) systems \cite{hu2015transfer,lin2021deep,sancinetti2022transfer} or self-supervised speech models (SSL) as feature extractors, followed by lightweight models trained on labeled L2 data for mispronunciation detection \cite{xu2021explore,vidal2023mispronunciation} or phoneme sequence prediction \cite{peng2021study}. Other strategies combine limited annotations with pseudo-labeling \cite{yang2022improving} or data augmentation \cite{fu2021dataaug,korzekwa2022computer}. While effective, these approaches still rely on a significant amount of fine-grained phoneme-level annotations. 

Unsupervised methods, which eliminate the requirement of expensive labeled training data, have also been explored for this task. Such approaches typically rely on  acoustic models trained on native data. The most widely used example of this family is the Goodness of Pronunciation (GOP) algorithm \cite{witt2000phone}, which estimates phoneme-level scores from posterior probabilities produced by native ASR models. Variants include DNN-based GOP \cite{hu2015improved}, alignment-free CTC formulations \cite{cao2024framework}, and SSL-based extensions such as MixGoP \cite{choi2025leveraging}. Other unsupervised approaches include articulatory Weighted-Dynamic Time Warping (W-DTW) \cite{sini2023phone} and extended recognition networks (ERNs) built from error patterns discovered without labeled data \cite{lee2015mispronunciation,lee2016personalized}. Unfortunately, unsupervised approaches typically underperform when compared with supervised ones~\cite{chen2016computer}.

In this work, we focus on  weakly-supervised scenarios where a large number of word- or utterance-level labels and only a limited number of phoneme-level labels are available for training, and explore various approaches to leverage this data effectively for learning phoneme-level scoring models.
An early indication that higher-level labels carry useful information for phoneme-level scoring can be found in \cite{cincarek2009automatic}, which showed that phoneme-level mispronunciation information could be recovered from word-level scores by modeling word error rates as  functions of the mispronunciation probabilities of their constituent phonemes. 
Further, some recent DNN-based supervised approaches jointly model multiple linguistic levels, either in parallel \cite{gong2022transformer} or hierarchically \cite{do2023hierarchical} showing that joint modeling benefits all levels. 
Here, we propose a variation of the architecture in \cite{gong2022transformer} that allows us to learn a phoneme-level model using only higher-level labels.  We also consider a two-stage annotation scenario, in which a few carefully-selected samples are further labeled at phoneme-level and used for finetuning. We show that our proposed architecture and selection process allow for an effective use of high-level or limited fine-grained labels, leading to comparable results to those obtained with full phoneme-level supervision.

\section{Methods}
\label{sec:methods}
As training and evaluation data we use the Speechocean762 database \cite{zhang2021speechocean762}, which provides pronunciation scores annotated at  phoneme,  word, and  utterance level. These multi-granularity annotations allow us to study weak supervision scenarios in which phoneme-level predictors are learned from higher-level labels.

\subsection{Baselines: GOP, GOP features + SVR and GOPT}
The Goodness of Pronunciation (GOP) algorithm provides an unsupervised measure of how closely an observed phoneme matches its canonical pronunciation according to an acoustic model. For a target phoneme $p$ aligned to frames $t_s$ through $t_e$, the basic GOP score is computed as the average log phone posterior (LPP) for phone $p$  within that time range given the observation: \(\text{GOP}(p)=\frac{1}{t_e-t_s+1}\sum_{t=t_s}^{t_e}\log P_t(p\mid\mathbf{O})\). Where $P_t(p\mid\mathbf{O})$ is obtained from a native acoustic model by summing state posteriors associated with $p$ at time $t$. 

The GOP method also forms the basis for a variety of supervised systems for which the LPPs are used to construct 2K-dimensional  feature vectors  for each phoneme in the transcription, where $K$ is the number of phonemes in the inventory of the recognizer used to compute the phoneme posteriors. The first $K$ GOP features correspond to the LPP for all $K$ phonemes in the inventory and the last $K$ are  
given by the LPP of the target phoneme minus the LPP of every phoneme  \cite{hu2015transfer}. 
The resulting vectors, usually called GOP features, can then be used as input to a support vector regressor (SVR), as in \cite{zhang2021speechocean762}, or as input to a Transformer, as in the GOPT approach \cite{gong2022transformer}.

In the GOPT method, phoneme-level GOP features are projected to a shared 24-dimensional space and combined with projected canonical phoneme embeddings and positional embeddings. The resulting sequence is processed by a Transformer encoder with prepended trainable \texttt{[CLS]} tokens, which are used as input to a regression head for each utterance-level score in the speechocean762 dataset. Further, regression heads are also attached to phoneme-level activations to predict phoneme-level and word-level scores. Notably, word-level heads are trained to predict the word-level label, repeated over each constituent phoneme. For inference, the phoneme-level scores from the word-level head are averaged to obtain the word-level score. For training, the mean squared error (MSE) losses from all levels are summed  and optimized jointly. The blocks inside the dashed line in Figure \ref{fig:gopt} show a simplified schematic of the original GOPT architecture. The GOPT codebase is available in github.\footnote{GOPT code: \url{https://github.com/YuanGongND/gopt}} 

\subsection{Proposed GOPT variant}
\label{sec:wgopt}
\begin{figure}[t]
    \centering
\includegraphics[width=\columnwidth]{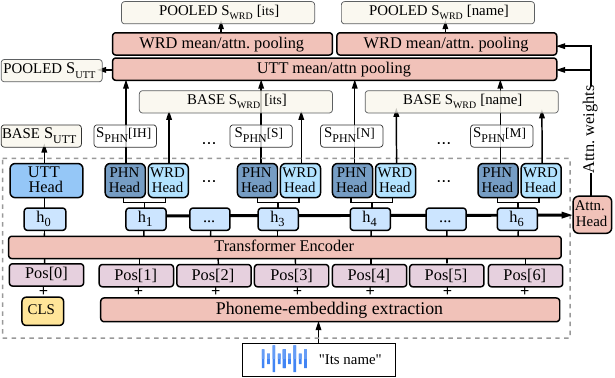}
\caption{Proposed architecture building on the GOPT backbone (showing only heads corresponding to the accuracy scores, for simplicity). Audio and transcriptions are processed to obtain phoneme-level embeddings. A transformer encoder is then applied to this sequence, after appending a \texttt{[CLS]} token and adding positional embeddings. In the original GOPT arquitecture, the transformer output corresponding to the position of the \texttt{[CLS]} token is passed through a linear head to produce the utterance-level score, and the hidden states corresponding to the phoneme positions are fed into word- and phoneme-level heads. In our proposed architecture, higher-level scores are obtained by pooling the phoneme-level scores using either mean or attention-weighted pooling.}
    \label{fig:gopt} 
\end{figure}

We propose a modification to the GOPT architecture that allows us to train phoneme-level score predictor heads using only word- or utterance-level labels. 
Given phoneme-level score predictors, we obtain an utterance-level score by aggregating information across all phonemes in the utterance instead of using the \texttt{[CLS]} tokens followed by a regression layer, as in the original architecture. The same procedure is applied at the word level to obtain a word-level score by aggregating across the phonemes belonging to each word.
We compare three strategies:
\begin{itemize}
\item \textbf{BASE}: the original GOPT strategy, where the utterance-level score is predicted directly from a dedicated \texttt{[CLS]} token, and word-level heads are trained to predict word-level labels propagated to the phonemes.
\item \textbf{MEAN}: utterance- and word-level scores are computed as the average of the phoneme-level scores within that unit.
\item \textbf{ATTN}: as for MEAN, but using a weighted average, with weights given by an attention head that takes the hidden states from the transformer for the corresponding unit as input.  
\end{itemize}
Figure \ref{fig:gopt} shows the original GOPT architecture (inside the dashed block) and, above it, the proposed modification.

In all cases, we vary the supervision regime by controlling which loss terms are active during training: only utterance-level labels, only word-level labels, only phoneme-level labels, or the sum of losses from all levels, corresponding to the original multi-task GOPT formulation. 

Note that, when the BASE approach is used for predicting the utterance scores, and the word- and phone-level losses are not included, the phoneme- and word-level regression heads are not trained. That is, in the original GOPT architecture, disabling the phoneme-level loss implies that phoneme-level scores cannot be computed. On the other hand, when using the MEAN or ATTN approaches for computing the utterance-level scores, the phoneme-level heads are trained, regardless of whether the phoneme-level loss is active or not.

\section{Experimental setup}
\label{sec:exp}
We conduct experiments on the Speechocean762 corpus, which contains 5000 English read utterances from 250 Mandarin L1 speakers (125 adults and 125 children). Each utterance is annotated by five raters at phoneme, word, and utterance level. The scores for each unit are averaged across raters, and utterance- and word-level accuracy scores (0–10) are divided by 5 to match the phoneme-level range (0–2). Score types include accuracy, stress, and total for the word level, and accuracy, fluency, completeness, prosody, and total for utterance level. At phoneme level, only accuracy labels are available. Average phoneme-level accuracy is strongly correlated with both word-level (PCC=0.91) and utterance-level accuracy (PCC=0.80). In this work, our goal is to predict phoneme-level accuracy scores with models trained only or mostly with word- or utterance-level accuracy scores. 

The corpus provides predefined training and test partitions and a HuggingFace loader.\footnote{Speechocean762 HF loader: \url{https://huggingface.co/datasets/mispeech/speechocean762}}
We keep the official training split and further divide the test split into a development set (1260 utterances) and a held-out test set (1240 utterances). The splits are speaker-disjoint and have similar score and age distributions.

We compute forced alignments and frame-level posteriors using the Kaldi Speechocean762 GOP recipe\footnote{Kaldi GOP recipe: \url{https://github.com/kaldi-asr/kaldi/tree/master/egs/gop_speechocean762}} with a TDNN-F acoustic model \cite{povey2018semi}  trained on LibriSpeech \cite{panayotov2015librispeech}. The phoneme inventory is the Kaldi monophone set, with size $K=42$. 
From these posteriors we compute phoneme-level GOP scores and the GOP feature vectors, using the GOPT codebase.
As an unsupervised baseline, we report performance using the raw GOP values. 
We also train phoneme-specific support vector regressors (SVRs) to predict phoneme-level  scores from GOP feature vectors using an RBF kernel and the default \texttt{scikit-learn} \cite{scikit-learn} hyperparameters. Finally, we train the original GOPT model using the released code and hyperparameters (lr $=10^{-3}$, 100 epochs, batch size 25), and our proposed variants.\footnote{Code available at \url{hidden/to/preserve/anonimity}} 
For the second stage in the two-stage process, we train or finetune for 60 and 30 epochs respectively. 
We show results using only accuracy scores for training since we found no advantage in performance on accuracy score prediction from adding the other word- and utterance-level score types during training.

\subsection{Evaluation metrics}
We evaluate performance using Mean Squared Error (MSE) and Pearson Correlation Coefficient (PCC) at the utterance, word, and phoneme levels. For transformer-based models, we run each configuration with five random seeds to assess the impact of the seed in the results, as in the original GOPT paper~\cite{gong2022transformer}. This procedure, though, does not consider the impact of the test data. To quantify this impact, for each seed, we compute 1000 bootstrap resamples of the evaluation set and obtain the MSE/PCC value for each set.\footnote{Confidence intervals are computed using \url{https://github.com/luferrer/ConfidenceIntervals}} The sampling is done by speaker to properly consider the correlation between their samples. Finally, for a given configuration, we pool the MSE/PCC values over the bootstrap samples across all seeds and  compute the mean and the 2.5 and 97.5 percentile over those values. The reported confidence intervals correspond to the largest distance between the mean and those two percentiles. For SVR and GOP, no seeds are used, so intervals are computed using 1000 bootstrap samples. For experiments that involve random subset selection for phoneme-level labeling, we generate five independent subset draws per set size. In those cases, the metrics from all draws, seeds and bootstrap samples are pooled to obtained the mean and confidence interval. 
\section{Results}
\label{sec:results}
We train the three architectures depicted in Figure~\ref{fig:gopt} (BASE, MEAN and ATTN) under five supervision regimes: utterance+word+phoneme (UWP), phoneme-only (P), word-only (W), utterance+word (UW), and utterance-only (U).  In each regime, we report performance of all available scores. At the phoneme level, the scores may be obtained with prediction heads trained with direct phoneme-level supervision or with weak supervision using higher-level labels through MEAN/ATTN pooling. Note that, in the BASE architecture, when phoneme labels are not available, phoneme-level heads are not trained.

Table~\ref{tab:main_res} reports development-set results for various combinations of supervision regime and architecture. 
All methods, including those with only utterance-level supervision, reach significantly better performance than the GOP baseline, which has a PCC of 0.34{ $\pm$ 0.04}, showing that the proposed architecture provides a viable mechanism for inducing phoneme-level structure from higher-level labels.
Naturally, phoneme-level PCC gets better as supervision gets closer to the phoneme level, with word-level supervision performing best among weakly supervised settings, followed by utterance+word (UW), and then utterance (U). 

\begin{table}[t]
  \centering
  \caption{Mean and confidence interval size on the development set for utterance/word-level PCC and phoneme-level PCC and MSE for the baseline GOPT (BASE) and proposed variants (MEAN and ATTN) under different supervision regimes: utterance+word+phoneme (UWP), phoneme-only (P), word-only (W), utterance+word (UW), and utterance-only (U). Missing values correspond to architectures that have no predictors trained at that level. $\uparrow/\downarrow$ indicate higher/lower is better.}
  \vspace{-1.5pt}
  \label{tab:main_res}
  \setlength{\tabcolsep}{5pt}
  \renewcommand{\arraystretch}{0.99}
  \footnotesize
  \begin{tabular}{llcccc}
    \toprule
     &  & Utterance & Word & \multicolumn{2}{c}{Phoneme} \\
    \cmidrule(lr){3-3} \cmidrule(lr){4-4} \cmidrule(lr){5-6}
    Labels & Model & PCC$\uparrow$ & PCC$\uparrow$ & PCC$\uparrow$ & MSE$\downarrow$ \\
    \midrule
   \multirow{3}{*}{\shortstack[l]{UWP}}
    & BASE & 0.71 {\tiny $\pm$ 0.10} & 0.53 {\tiny $\pm$ 0.10} & 0.61 {\tiny $\pm$ 0.08} & 0.09 {\tiny $\pm$ 0.02}\\
    & MEAN & 0.66 {\tiny $\pm$ 0.11} & 0.55 {\tiny $\pm$ 0.10} & 0.58 {\tiny $\pm$ 0.08} & 0.09 {\tiny $\pm$ 0.02} \\
    & ATTN & 0.69 {\tiny $\pm$ 0.09} & 0.58 {\tiny $\pm$ 0.09} & 0.59 {\tiny $\pm$ 0.08} & 0.09 {\tiny $\pm$ 0.02} \\
    \midrule
    P
    & BASE & - & - & 0.61 {\tiny $\pm$ 0.08} & 0.09 {\tiny $\pm$ 0.02} \\
    \midrule
   \multirow{3}{*}{W}
    & BASE & - & 0.52 {\tiny $\pm$ 0.11} & - & - \\
    & MEAN & - & 0.56 {\tiny $\pm$ 0.10} & 0.54 {\tiny $\pm$ 0.08} & 0.10 {\tiny $\pm$ 0.03} \\
    & ATTN & - & 0.59 {\tiny $\pm$ 0.10} & 0.56 {\tiny $\pm$ 0.09} & 0.10 {\tiny $\pm$ 0.03} \\
    \midrule
   \multirow{3}{*}{\shortstack[c]{UW}}
    & BASE & 0.71 {\tiny $\pm$ 0.09} & 0.51 {\tiny $\pm$ 0.10} & - & - \\
    & MEAN & 0.68 {\tiny $\pm$ 0.10} & 0.54 {\tiny $\pm$ 0.09} & 0.50 {\tiny $\pm$ 0.08} & 0.22 {\tiny $\pm$ 0.05} \\
    & ATTN & 0.69 {\tiny $\pm$ 0.10} & 0.54 {\tiny $\pm$ 0.08} & 0.53 {\tiny $\pm$ 0.08} & 0.10 {\tiny $\pm$ 0.03} \\
    \midrule
   \multirow{3}{*}{U}
    & BASE & 0.71 {\tiny $\pm$ 0.09} & - & - & - \\
    & MEAN & 0.71 {\tiny $\pm$ 0.09} & - & 0.46 {\tiny $\pm$ 0.06} & 0.27 {\tiny $\pm$ 0.05} \\
    & ATTN & 0.71 {\tiny $\pm$ 0.09} & - & 0.46 {\tiny $\pm$ 0.06} & 0.23 {\tiny $\pm$ 0.04} \\
    \bottomrule
  \end{tabular}
  \vspace{-0.3cm}
\end{table}

In addition, Table \ref{tab:main_res} shows that, beyond enabling the generation of phoneme-level scores without phoneme-level supervision, the proposed architecture also provides a gain over the baseline at word level. Specifically, the ATTN variant improves word-level PCC relative to the BASE approach under all supervision regimes (UWP, W and UW). At the utterance level, the ATTN variant leads to similar results as the BASE approach.
Finally, at phoneme-level the ATTN architecture leads to the best PCC values under weak supervision regimes.
Based on these results, we use ATTN pooling for all subsequent experiments. 

Notably, phoneme-level PCC and MSE do not correlate well. In some cases, models achieve similar phoneme-level PCC but substantially different MSE (see the MEAN vs ATTN results in the UW block): their scores result in a similar ranking while differing in range. Suboptimality of the MSE could potentially be fixed through a post-hoc calibration transformation of the scores, learned on held-out data. For the remaining results, we use PCC as performance metric, leaving the calibration problem for future work. 

Next, we assume an active learning scenario where labels are obtained and training is performed in two stages. In the first stage, a system is trained using a relatively large number of utterance-level labels (in our experiments, N $=2500$), as in the utterance-only experiments above. We call this model \mbox{1S-U}. This system is then used to determine which samples should be labeled at the word or phoneme level, given a limited annotation budget. To this end, all training samples from the first stage are run through this system and the absolute error (AE) values for the utterance-level accuracy score are computed. We then select the $n$ utterances with the smallest AE values (which we call \emph{best}) and compare this approach with random selection. For both strategies, we consider an unbalanced variant that applies selection over the full utterance pool and a balanced variant that splits utterances into $B$ equal-width bins based on ground truth utterance-level scores and selects approximately $n/B$ items per bin randomly within bins for the random-balanced case and by ascending error within bins for the error-balanced case. Finally, we compare training from scratch using only the $n$ selected samples versus starting from the utterance-only model and finetuning with the selected data. 

\begin{figure}[t]
    \includegraphics[width=\columnwidth]{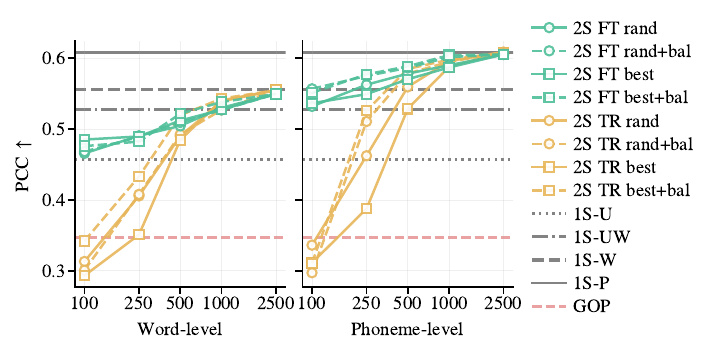}
    \vspace{-6mm}
    \caption{Phoneme-level PCC for the proposed two-stage training process using a varying number of samples (x-axis) labeled at word-level (left) or phoneme-level (right) for the second stage. Curves differ on training strategy (FT: finetuning the 1S-U model, TR: training from scratch), and sample selection strategy (rand/best, balanced or not).  Horizontal lines show the one-stage baselines with different supervision regimes (1S-U, 1S-W, 1S-UW, 1S-P) from Table \ref{tab:main_res} and the GOP baseline.}
    \label{fig:lc}
\end{figure}

Figure~\ref{fig:lc} shows phoneme-level PCC for the 1S-U ATTN model trained or finetuned with word-level labels (left) and phoneme-level labels (right), as a function of the number of selected utterances, $n$ (in log scale). We also include results for several reference systems: the GOP baseline, and the ATTN model trained with different supervision regimes. Across both panels, the two-stage approaches with finetuning (2S FT) consistently outperform the GOP and the 1S-U systems, even at very small annotation budgets. Training from scratch (2S TR) lags behind finetuning, showing the benefit of starting from a system trained with weak supervision, particularly for small annotation budgets.

Selecting samples for fine-grained annotations to be balanced across the range of utterance-level ground truth scores  (dashed lines) is better (or not worse) than selecting without balancing (solid lines).
No consistent difference is found between random selection and selection based on AE values. For this reason and given its simplicity, we select the rand+bal approach for the experiments on the test data.

\begin{figure}[t]
\includegraphics[width=0.9\columnwidth]{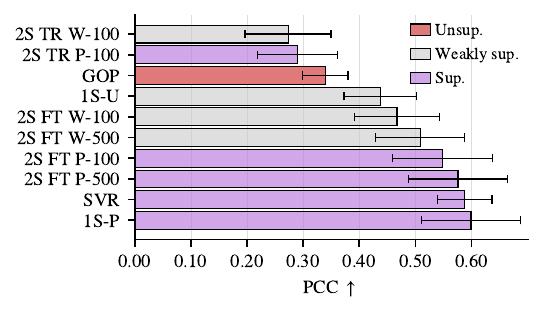}
\vspace{-3mm}
\caption{Phoneme-level PCC on the test set by supervision type: unsupervised (red), weakly supervised (orange), and supervised (green). Methods with a "W/P-N" suffix use only $N$ utterances labeled at level word or phoneme level in the second stage. Error bars show 95\% confidence intervals.}
 \label{fig:phnexp}
\end{figure}

Finally, Figure~\ref{fig:phnexp} reports phoneme-level PCC on the held-out test set, where no development decisions were made, including unsupervised (red), weakly supervised (grey), and supervised (violet) methods. The methods are ordered from lower to higher PCC, with confidence intervals shown for each bar. As above, the GOP and 1S-P  systems are taken as the unsupervised and fully supervised references. The figure highlights a set of practical scenarios.  If a relatively large number of utterances (N$=2500$) can be labeled at the utterance level, the proposed ATTN architecture trained with utterance-level supervision (\mbox{1S-U}) provides an improvement over GOP. If only a very small number of utterances can be labeled at word or phoneme level, training from scratch on those labels (2S TR W/P-100) is not a reliable alternative to GOP in this setting. In contrast, the same limited fine-grained budget becomes useful when combined with utterance supervision in a two-stage approach where the 1S-U model is finetuned on those labels (2S FT W/P-100). Notably, the two-stage system finetuned with 500 utterances labeled at phoneme-level reaches a performance within 5\% of the performance of the system trained with 2500 utterance labeled at phoneme-level, reducing by a factor of 5 the need for fine-grained annotations.

Notably, the simple SVR approach has a performance similar to that of the much more complex 1S-P system, with narrower confidence intervals. This suggests that a simpler architecture than the ones commonly used for this task -- among which the GOPT one is only an example -- may be sufficient, given the data limitations. We leave this exploration for future work. 

To conclude, Figure \ref{fig:phnexp} shows that, when using our proposed methods, several weakly-supervised  models lead to a competitive performance compared to the 1S-P topline, which requires all training utterances to be labeled at phoneme level. 
Specifically, a PCC within 10\% of the PCC of the topline can be achieved when only the following annotations are available:
\begin{itemize}
\item Word-level labels for all training data (1S-W),
\item Utterance-level labels for all training data and phoneme-level labels for as low as 100 utterances (2S FT P-100/500),
\item Utterance-level labels for all training data and word-level labels for as low as 500 utterances (2S FT W-500).
\end{itemize}
While there is still a trade-off between annotation effort and performance, our proposed methods greatly reduce the gap between models trained with full phoneme-level supervision and model trained with fewer or higher-level labels, compared to naively using the annotated data to train a model from scratch.


\section{Discussion and conclusions}
\label{sec:}
In this work, we explore whether phoneme-level pronunciation information can be recovered using only higher-level supervision. We propose a variant of the GOPT architecture that allows for higher-level supervision to flow through to phoneme-level prediction heads by computing higher-level predictions as attention-pooled phoneme-level predictions. We also propose a two-stage approach where a model trained only with utterance-level labels is finetuned with a small amount of word- or phoneme-level labeled data. Results show that our approach results in an efficient use of the available annotations, reaching competitive results compared to the model trained with full supervision, using only a fraction of data annotated at fine-grained level, allowing for a significant reduction in annotation costs with a marginal in performance.
\bibliographystyle{IEEEtran}
\bibliography{mybib}

\end{document}